\documentclass[aps,preprint,tightenlines,floatfix,groupedaddress,nofootinbib]{revtex4}

\usepackage{graphicx}
\usepackage{dcolumn}
\def\beq{\begin{equation}}
\def\eeq{\end{equation}}
\def\bea{\begin{eqnarray}}
\def\eea{\end{eqnarray}}
\def\nn{\nonumber}
\def\sss{\scriptstyle}
\def\lft{{\scriptstyle L}}
\def\rht{{\scriptstyle R}}
\def\roughly#1{\mathrel{\raise.3ex\hbox
{$#1$\kern-.75em\lower1ex\hbox{$\sim$}}}}
\def\lsim{\roughly<}

\begin{document}
\bibliographystyle{apsrev}

\preprint{\vbox {\hbox{UdeM-GPP-TH-06-147} \hbox{hep-ph/0605123}}}

\vspace*{2cm}

\title{\boldmath CP Violation in Supersymmetric Theories:
${\tilde t}_2 \to {\tilde t}_1 \tau^- \tau^+$}

\author{Ken Kiers}
\email{knkiers@taylor.edu}
\altaffiliation{Permanent Address:
Physics Department, Taylor University, 236 West Reade Ave., Upland,
Indiana, 46989, USA}
\author{Alejandro Szynkman}
\email{szynkman@lps.umontreal.ca}
\author{David London}
\email{london@lps.umontreal.ca}
\affiliation{\it Physique des Particules, Universit\'e
de Montr\'eal, C.P. 6128, succ. centre-ville, Montr\'eal, QC,
Canada H3C 3J7}

\date{\today}

\begin{abstract}
Supersymmetric (SUSY) theories include many new parameters, some of
which are CP-violating. Assuming that SUSY is found at a future
high-energy collider, we examine the decay ${\tilde t}_2 \to {\tilde
t}_1 \tau^- \tau^+$ with an eye to obtaining information about the new
CP phases. We show that there are two CP-violating asymmetries which
can be large in some regions of the SUSY parameter space. These
involve measuring one or both of the $\tau$ spins. These asymmetries
are particularly sensitive to $\phi_{\sss A_t}$, the phase of the
trilinear coupling $A_t$.
\end{abstract}


\maketitle

\newpage
\setcounter{page}{1}

\section{Introduction}

\label{sec:introduction}

Supersymmetry (SUSY) is usually invoked to solve the hierarchy problem
of the Standard Model (SM). Because such models are so theoretically
compelling, many believe that SUSY must be found in nature, and much
work has gone into elucidating various aspects of such theories.

In SUSY theories each ordinary fermion and gauge boson has a
superpartner, respectively of spin 0 and spin $\frac12$. These models
contain a large number of unknown parameters. Many of these parameters
can be complex, so that, in general, SUSY theories predict large
CP-violating effects due to the presence of SUSY particles. The
purpose of this paper is to begin a systematic high-energy exploration
of CP violation in SUSY couplings. That is, throughout this paper we
assume that the CP-conserving parameters are known, but that those
which violate CP remain to be measured. Of course, most measurements
will not probe single CP-violating SUSY phases. However, if one makes
many measurements, then the combined effect will be to measure or
constrain all SUSY phases.

The low-energy CP-violating properties of the couplings of
superparticles have been explored. For example, constraints due to
electric dipole moments (EDMs) have been studied in detail
\cite{gabbiani,grossman,chang,pilaftsis,abel,demir,olive}.  Contributions to
meson mixing have also been investigated~\cite{mesonmix}, and the
effect of SUSY theories on CP violation in the $B$-meson system has
been examined~\cite{AliLon}.

The EDM experiments are particularly constraining and have led to the
so-called SUSY CP problem -- if sfermion masses are of order the weak
scale and complex SUSY parameters have phases of order unity,
calculations of EDMs generically yield values in excess of the
experimental limits.  Furthermore, the experimental limits are
expected to be improved substantially over the next few years, which
will lead to further constraints on the SUSY parameter
space~\cite{olive}.  Theoretical calculations of EDMs are highly
model-dependent, and there are several scenarios in which the problem
is ameliorated~\cite{scenarios}.  For example, one simple way to avoid
the SUSY CP problem is to assume that sfermion masses are of order the
weak scale, but that all relevant CP-odd phases are highly suppressed.
Another is to adopt the opposite stance, taking the phases to be of
order unity, but sfermion masses to be very large (of order several
TeV).  Scenarios have also been studied in which only the first- and
second-generation sfermions are very heavy, but the third-generation
sfermions are near the electroweak scale.  In other cases, authors
have made assumptions concerning universality or alignment of various
soft-SUSY-breaking terms.  Finally, in principle there could be
substantial cancellations between different diagrams, leading to a
reduction in the EDMs.

Assuming that SUSY particles are found, it will be necessary to
measure the CP-violating properties of their couplings in order to get
a complete picture of the physical SUSY model.  Several studies have
already been performed, showing how high-energy observables could be
used to measure or constrain various SUSY
phases~\cite{high_energy,ValWang}.  In the present work we consider
CP-odd asymmetries related to a particular decay mode of the heavier
top squark in SUSY.  

The observables that we consider are sensitive primarily to parameters
associated with the third-generation squarks.  EDMs can in principle
constrain quantities associated with the third generation.  (For
example, the two-loop Bar-Zee-type diagrams considered in
Ref.~\cite{chang} can directly constrain the CP-violating parameters
related to the third generation squarks.)  We will assume that the
SUSY parameters that we use, as well as those not directly involved in
our calculation (i.e., those corresponding to the first two
generations), are such that the various EDM constraints are not
violated.  Our point of view is that our observables may be used to
provide an {\it independent} measurement of relevant CP-odd SUSY
parameters.  Such measurements would be complementary to the
constraints coming from low-energy observables such as EDMs or meson
mixing.  If indeed the EDM constraints end up predicting particular
values for the parameters we consider, then a high-energy measurement
should verify such predictions.

In SUSY theories, there are two Higgs doublets. As a result, there are
three neutral Higgs bosons -- two scalars and one pseudoscalar. If CP
is violated, when one evolves from the gauge basis to the mass basis,
these neutral particles can mix, so that the couplings of the physical
Higgs bosons are generally mixtures of scalar and pseudoscalar.
Recently, it was shown that this mixing could be probed in the process
$H^0 \to t{\bar t}$ \cite{ValWang}. The measurement of such a mixing
would be one signal of CP violation in SUSY theories. The authors of
Ref.~\cite{ValWang} note that this signal could also be seen in $H^0
\to \tau{\bar \tau}$ if the $H^0$ is too light to decay to $t{\bar
t}$. As such, this method, and ours described below, are applicable to
high-energy colliders such as the Large Hadron Collider or Next Linear
Collider.

SUSY theories also contain two scalar superpartners of the top quark,
one for each $t$-quark helicity, called ${\tilde t}_\lft$ and ${\tilde
t}_\rht$. These two ``stops'' can mix, resulting in two mass
eigenstates ${\tilde t}_1$ and ${\tilde t}_2$ whose masses can be very
different. Here we adopt the standard notation $m_{{\tilde t}_2} >
m_{{\tilde t}_1}$. In this paper we examine the decay process ${\tilde
t}_2 \to {\tilde t}_1 \tau^- \tau^+$. This comes about principally
through the exchange of an intermediate $Z$ boson or any of the Higgs
bosons. The measurement of CP violation in this process will therefore
probe the CP phases in the stop couplings, as well as
scalar-pseudoscalar mixing in the Higgs sector.

There are a variety of CP-violating asymmetries in this process, some
of which depend on strong (CP-conserving) phases. However, since the
$\tau$'s are leptons, they cannot emit gluons, and the process
${\tilde t}_2 \to {\tilde t}_1 \tau^- \tau^+$ does not have
QCD-induced strong phases (any gluons exchanged between the two stops
only serve to renormalize the stop couplings). In order to generate
strong phases, we therefore include the widths of the mediating
particles.

In order to calculate the rate and the various CP-violating
asymmetries for ${\tilde t}_2 \to {\tilde t}_1 \tau^- \tau^+$, various
SUSY parameters must be specified: the masses and couplings of
$m_{{\tilde t}_1}$ and $m_{{\tilde t}_2}$, and the masses, mixings and
widths of the Higgs bosons.  However, these parameters are not all
independent -- they can be computed from the underlying SUSY
parameters. To be specific, there are 7 fundamental SUSY parameters
which strongly affect this decay process (see Sec.~\ref{sec:results}
for an explicit list).

However, since the aim here is to measure CP violation, we are mainly
interested in the underlying CP-violating SUSY parameters. We note
that all CP-violating asymmetries in ${\tilde t}_2 \to {\tilde t}_1
\tau^- \tau^+$ depend only on the stop couplings, the mixings of the
Higgses, and their widths. These quantities in turn depend on several
CP-violating (i.e., complex) parameters in the underlying theory --
the various trilinear couplings, the gaugino masses and $\mu$.  Thus,
by specifying the phases of the various complex parameters, one can
predict all CP-violating effects in ${\tilde t}_2 \to {\tilde t}_1
\tau^- \tau^+$ (recall that we have assumed that all CP-conserving
quantities are known). We shall assume throughout that $\mu$ is real and
positive.  This may be arranged by rephasing the various complex
parameters~\cite{abel}.  With this phase convention, the asymmetries in
${\tilde t}_2 \to {\tilde t}_1 \tau^- \tau^+$ have a strong dependence
on $\phi_{\sss A_t}$ (the phase of the trilinear coupling $A_t$) and
only a much weaker dependence on the other phases.  (See
Sec.~\ref{sec:results} for a more detailed discussion of this point.)
Thus, the measurement of CP-violating effects in ${\tilde t}_2 \to
{\tilde t}_1 \tau^- \tau^+$ will allow us to measure/constrain
$\phi_{\sss A_t}$.

A computer program -- ``{\tt CPsuperH}'' -- has been written by others
to generate the physical SUSY parameters (including the masses,
mixings and widths of the Higgs bosons, the masses of the stops and
the couplings of the Higgs bosons to the (s)fermions) from the
underlying SUSY parameters~\cite{CPsuperH}. Thus, with this program in
hand, it is necessary only to specify the values of the fundamental
SUSY parameters. We use {\tt CPsuperH} extensively in our numerical
work below and have adopted their notation for the coupling constants.

In Sec.~\ref{sec:amplitude}, we compute the amplitude for ${\tilde
t}_2 \to {\tilde t}_1 \tau^- \tau^+$, as well as its CP-conjugate. The
various CP-violating signals are examined in
Sec.~\ref{sec:asymmetries}.  We show that there are two CP asymmetries
which can be large in some regions of SUSY parameter space. These
involve the measurement of one or both of the $\tau$ spins. In
Sec.~\ref{sec:results}, we present the predictions for these
asymmetries.  The measurement of these asymmetries will provide
information about the SUSY CP phases. We conclude in
Sec.~\ref{sec:conclusions}. The Appendix contains additional
information about the various CP asymmetries.

\section{\boldmath Amplitude for ${\tilde t}_2 \to {\tilde t}_1 \tau^-
\tau^+$}

\label{sec:amplitude}

We now present the details of the ${\tilde t}_2 \to {\tilde t}_1
\tau^- \tau^+$ calculation. There are box-diagram contributions to
this process, but they are expected to be small. Thus, this decay is
mediated principally by the $Z$ boson and by any of the three neutral
Higgs bosons. That is, we have $\tilde{t}_2\to \tilde{t}_1(Z,H_i)$
followed by $(Z,H_i)\to\tau^-\tau^+$. In the following we always take
the $\tau^-$-$\tau^+$ invariant mass to be above the $Z$ pole, and so
we assume that only neutral Higgs-exchange diagrams are important.
While interferences between the Higgs diagrams and the off-shell $Z$
could in principle give contributions to the CP asymmetries we
consider, these contributions are expected to be small, unless the
lightest Higgs boson has a mass very close to that of the $Z$.  To
simplify our discussion, we ignore any contributions from the
$Z$.\footnote{In order that this approximation be reasonable, we will
always consider cases in which the lightest Higgs is relatively
well-separated in mass from the $Z$.  If $Z$ contributions end up
being required, they could be incorporated in a straightforward
manner.}

As mentioned in the introduction, the scalar and pseudoscalar Higgs
particles in SUSY can mix if CP is violated, leading to three physical
neutral Higgs bosons that do not have well-defined CP transformation
properties. In general, the scalar-pseudoscalar mixing need not be
small. The squarks also mix, and in general the up-type squark
mass-squared matrix is $6\times 6$. However, it is usually assumed
that the up-type squarks are block-diagonalized by the same unitary
transformations that diagonalize the up-type quarks. In this case, one
obtains a simplified form for the mixing matrix: one only needs to
diagonalize a set of $2\times 2$ matrices. We adopt this simplifying
assumption for the stops and consider only the mixing between
$\tilde{t}_\lft$ and $\tilde{t}_\rht$. The {\tt CPsuperH} program
calculates the masses and mixings of both the stops and the Higgs
bosons.

The couplings between the stops and Higgs bosons may be described by
the following Lagrangian:
\beq
    {\cal L}_{\sss H\tilde{t}\tilde{t}} = v \sum_{i,j,k}
    g_{\sss H_i\tilde{t}_j^*\tilde{t}_k} H_i \tilde{t}_j^* \tilde{t}_k ~,
\eeq
where $v$ is related to the vacuum expectation values of the two Higgs
doublets, $v=\sqrt{v_1^2+v_2^2}$, and where $i=1,2,3$ and $j,k=1,2$.
The coupling constants $g_{\sss H_i\tilde{t}_j^*\tilde{t}_k}$ are defined
as follows~\cite{CPsuperH},
\beq
   vg_{\sss H_i\tilde{t}_j^*\tilde{t}_k} = 
      \left(\Gamma^{\alpha\tilde{t}^*\tilde{t}}\right)_{\beta\gamma} 
      O_{\alpha i} U^{\tilde{t}*}_{\beta j} U^{\tilde{t}}_{\gamma k} \; ,
\eeq
where $O$ and $U^{\tilde{t}}$ are the Higgs and stop mixing matrices,
respectively.  The three $2\times 2$ matrices
$\Gamma^{\alpha\tilde{t}^*\tilde{t}}$ depend on the SUSY parameters
$A_t$, $\mu$, $\cos\beta$ and $\sin\beta$ (where $\tan\beta\equiv
v_2/v_1$).  Expressions for the matrices
$\Gamma^{\alpha\tilde{t}^*\tilde{t}}$ may be found in Appendix B of
Ref.~\cite{CPsuperH}.  In the CP-invariant limit, the couplings
involving scalar Higgs bosons are real and those involving the
pseudoscalar Higgs are purely imaginary.  If CP is broken, these
couplings are in general complex.

We also require the couplings between the tau leptons and the various
Higgs bosons. The relevant piece of the Lagrangian is
\begin{eqnarray}
      {\cal L}_{\sss H{\overline \tau}\tau}=-g_\tau\sum_{i}H_i \overline{\tau}\left(
           g^S_{\sss H_i\overline{\tau}\tau}+ig^P_{\sss H_i\overline{\tau}\tau}\gamma^5\right)\tau \; ,
	\label{eq:Htautau}
\end{eqnarray}
where $i=1,2,3$. The constant $g_\tau$ and the scalar and pseudoscalar
coupling constants, $g^S_{\sss H_i\overline{\tau}\tau}$ and $g^P_{\sss
H_i\overline{\tau}\tau}$, respectively, are real. At tree-level the
couplings are given by the expressions~\cite{CPsuperH}
\beq
     g_\tau^{\mbox{\scriptsize tree}}= \frac{gm_\tau}{2m_W},~~~
     g_{\sss H_i\overline{\tau}\tau}^{S,\mbox{\scriptsize tree}}=\frac{O_{1i}}{\cos\beta},~~~
     g_{\sss H_i\overline{\tau}\tau}^{P,\mbox{\scriptsize tree}}=-O_{3i}\tan\beta \; .
      \label{eq:g_tree}
\eeq
The program {\tt CPsuperH} produces quantum-corrected versions of the
above expressions.  In the limit that CP is conserved, a given Higgs
boson $H_i$ has either a scalar or a pseudoscalar coupling, but not
both.

The final ingredient in our calculation is the Higgs-boson propagator.
In the decay ${\tilde t}_2 \to {\tilde t}_1 \tau^- \tau^+$, we allow
one or more of the Higgs bosons to go on-shell. The reason for this is
two-fold. First, this will dramatically increase the rate, giving a
possibility that experimentalists will in fact be able to measure the
asymmetries that we construct. Second, as noted above, the strong
phases in this process are obtained mathematically by including the
widths, i.e.\ absorptive pieces in the propagator. These absorptive
parts have a much larger effect near resonance.

There is one complication here. It is true that {\tt CPsuperH}
computes all the physical properties of the Higgs bosons. In
particular, the width of the $H_i$ boson is found by calculating the
1-loop contributions of the light particles (taken here to be
principally $b$ and $\tau$), and applying Cutkowsky rules. However, in
our calculation the ``off-diagonal widths'' come into play. That is,
this 1-loop diagram can also be used to relate a $H_i$ boson to a
$H_j$ boson, and the absorptive parts of such diagrams can be important
for those CP asymmetries in our decay process which rely on strong
phases. {\tt CPsuperH} does not give these off-diagonal widths, but
they have been computed in Ref.~\cite{widths}.\footnote{{\tt CPsuperH}
does include off-diagonal contributions to the Higgs self-energy in
its calculation of the Higgs boson masses \cite{thanks}.}  A good
approximation for the Higgs propagator (divided by `$i$') is
\begin{eqnarray}
      D(M^2) &=& \nn \\ 
         & & \hspace{-.6in}
           \left(\begin{array}{ccc}
	    M^2 - m_{\sss H_1}^2 + i \, {\rm Im}\widehat{\Pi}_{11} & 
	     i \, {\rm Im}\widehat{\Pi}_{12} & i \, {\rm Im}\widehat{\Pi}_{13} \\
            i \, {\rm Im}\widehat{\Pi}_{21} & 
	     M^2 - m_{\sss H_2}^2 + i \, {\rm Im}\widehat{\Pi}_{22} & 
	     i \, {\rm Im}\widehat{\Pi}_{23} \\
            i \, {\rm Im}\widehat{\Pi}_{31} & i \, {\rm Im}\widehat{\Pi}_{32} &
	     M^2 - m_{\sss H_3}^2 + i \, {\rm Im}\widehat{\Pi}_{33} \\
	    \end{array}\right)^{-1} ,
\end{eqnarray}
where $M^2=(p_1+p_2)^2$, with $p_1$ and $p_2$ being the four-momenta
of the $\tau^-$ and $\tau^+$, respectively. Explicit expressions for
the absorptive parts of the Higgs boson self-energies, ${\rm
Im}\widehat{\Pi}_{ij}(M^2)$, may be found in Ref.~\cite{widths}. For
our purposes, it is sufficient to consider only the contributions to
${\rm Im}\widehat{\Pi}_{ij}(M^2)$ arising from the $b$-$\overline{b}$
and $\tau^\pm$ loops\footnote{We restrict our consideration to cases
in which the neutral Higgs bosons can only decay into (non-$t$) quark
and lepton pairs. In principle, $M^2$ could still be large enough
that other terms should be included in the calculation of ${\rm
Im}\widehat{\Pi}_{ij}(M^2)$. In practice, however, we are only
interested in the effect of ${\rm Im}\widehat{\Pi}_{ij}(M^2)$ near
resonances, i.e., when $M^2\sim m_{\sss H_i}^2$. In such cases it is
sufficient to restrict our attention to the $b$ and $\tau$
contributions.}, so that
\begin{eqnarray}
      {\rm Im}\widehat{\Pi}_{ij}(M^2) \simeq {\rm Im}\widehat{\Pi}_{ij}^{bb}(M^2)
           + {\rm Im}\widehat{\Pi}_{ij}^{\tau\tau}(M^2)\; ,
\end{eqnarray}
where
\begin{eqnarray}
      {\rm Im}\widehat{\Pi}_{ij}^{bb}(M^2) & = & 3 
         \left(\frac{M^2g_b^2}{8\pi}\right)
         \left(1+5.67\; \frac{\alpha_s(M^2)}{\pi}\right)\nn \\
	 & & \hskip.4truein \times \left[
	   \left(1-4\kappa_{b}\right)
	   g^S_{\sss H_i\overline{b}b}g^S_{\sss H_j\overline{b}b}+
	   g^P_{\sss H_i\overline{b}b}g^P_{\sss H_j\overline{b}b}\right] 
	    \lambda^{1/2}(1,\kappa_{b},\kappa_{b}) \; ,\\
      {\rm Im}\widehat{\Pi}_{ij}^{\tau\tau}(M^2) & = & 
         \left(\frac{M^2g_\tau^2}{8\pi}\right) \left[
	   \left(1-4\kappa_{\tau}\right)
	   g^S_{\sss H_i\overline{\tau}\tau}g^S_{\sss H_j\overline{\tau}\tau}+
	   g^P_{\sss H_i\overline{\tau}\tau}g^P_{\sss H_j\overline{\tau}\tau}\right] 
	 \lambda^{1/2}\left(1,\kappa_{\tau},\kappa_{\tau}\right) \; ,
       \label{eq:ImPitautau}
\end{eqnarray}
with $\lambda(x,y,z)\equiv x^2+y^2+z^2-2(xy+xz+yz)$ and
$\kappa_{b,\tau}\equiv m_{b,\tau}^2/M^2$. The constants $g_b$ and
$g^{\sss S,P}_{\sss H_i\overline{b}b}$ are associated with the
$H_i\overline{b}b$ couplings and are defined in analogy with
Eq.~(\ref{eq:Htautau}).  (The tree-level expressions for the coupling
constants are identical to those given in Eq.~(\ref{eq:g_tree}), but
with the substitution $m_\tau\to m_b$.)  Note that the diagonal
elements are related to the widths of the Higgs bosons through the
optical theorem, ${\rm Im}\widehat{\Pi}_{ii}(m_{\sss H_i}^2) \simeq
\Gamma(H_i)m_{\sss H_i}$, giving rise to the usual Breit-Wigner form
of the propagator if one takes $M^2=m_{\sss H_i}^2$ in the
self-energies and ignores the off-diagonal terms.  We include the
off-diagonal terms in our calculation, since in some cases they have a
non-negligible effect on the asymmetries that we consider.

\begin{figure}[t]
\begin{center}
\resizebox{4in}{!}{\includegraphics*{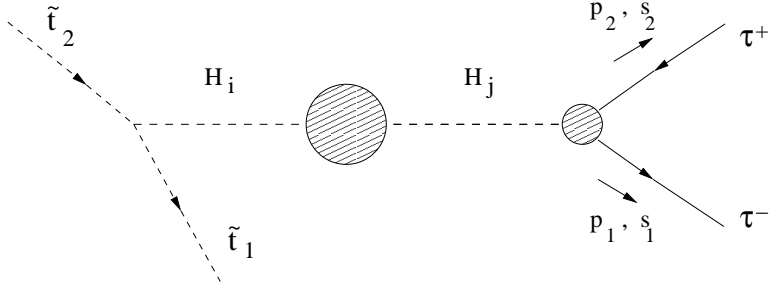}}
\caption{Feynman diagram for the decay $\tilde{t}_2\to \tilde{t}_1
\tau^-\tau^+$. Note that the Higgs propagator has off-diagonal terms,
so transitions $H_i\to H_j$ are allowed. The shaded circle at the
$H_j$-$\tau$-$\tau$ vertex denotes the fact that the effective
coupling calculated by {\tt CPsuperH} includes some loop corrections.
}
\label{fig:feyn_diag}
\end{center}
\end{figure}

Considering only the contributions due to the Higgs diagrams (see
Fig.~\ref{fig:feyn_diag}), the amplitude for the decay
$\tilde{t}_2^-\to \tilde{t}_1^- \tau^-\tau^+$ is given by
\begin{eqnarray}
      {\cal A}(s_1,s_2) = B\bar{u}(p_1,s_1)v(p_2,s_2)+C\bar{u}(p_1,s_1)\gamma^5v(p_2,s_2),
        \label{eq:amp}
\end{eqnarray}
where the indices `1' and `2' refer to $\tau^-$ and $\tau^+$,
respectively. The complex coefficients $B$ and $C$ are given by
\begin{eqnarray}
      B &=& v g_\tau\sum_{i,j}g_{\sss H_i\tilde{t}_2^*\tilde{t}_1}D_{ij}(M^2)
            g^S_{\sss H_j\overline{\tau}\tau} \; ,\\
      C &=& i v g_\tau\sum_{i,j}g_{\sss H_i\tilde{t}_2^*\tilde{t}_1}D_{ij}(M^2)
            g^P_{\sss H_j\overline{\tau}\tau} \; .
\end{eqnarray}
Calculation of the absolute value squared of the amplitude, for
specific spin states $s_1$ and $s_2$, yields the expression,
\begin{eqnarray}
      \left|{\cal A}(s_1,s_2)\right|^2 & = & \frac{1}{2}\left|B\right|^2\left[
	       \left(M^2-4m_\tau^2\right)\left(1-s_1\cdot s_2\right)+
	             2 \; p_1\cdot s_2 \; p_2\cdot s_1\right] \nn\\
               & & + \frac{1}{2}\left|C\right|^2\left[
	       M^2\left(1+s_1\cdot s_2\right)-
	             2 \; p_1\cdot s_2 \; p_2\cdot s_1\right] \label{eq:ampsquared} \nn\\
               & & -m_\tau\left(BC^*+B^*C\right)\left(p_1\cdot s_2 + p_2\cdot s_1\right) \nn \\
               & & + i\left(B^*C-BC^*\right)
                   \epsilon_{\alpha\beta\lambda\mu}p_1^\alpha s_1^\beta p_2^\lambda s_2^\mu \; , 
	\label{eq:Amp_sq}
\end{eqnarray}
in which we have adopted the convention $\epsilon_{0123}=+1$.

In order to calculate CP-violating asymmetries we must determine the
amplitude for the CP-conjugate process. This is
\begin{eqnarray}
      \overline{\cal A} = \overline{B}\bar{u}(p_1,s_1)v(p_2,s_2)+
            \overline{C}\bar{u}(p_1,s_1)\gamma^5v(p_2,s_2),
        \label{eq:amp_cp}
\end{eqnarray}
where
\begin{eqnarray}
      \overline{B} &=& v g_\tau\sum_{i,j}g_{\sss H_i\tilde{t}_2^*\tilde{t}_1}^*D_{ij}(M^2)
            g^S_{\sss H_j\overline{\tau}\tau} \; ,\\
      \overline{C} &=& -i v g_\tau\sum_{i,j}g_{\sss H_i\tilde{t}_2^*\tilde{t}_1}^*D_{ij}(M^2)
            g^P_{\sss H_j\overline{\tau}\tau} \; 
\end{eqnarray}
(recall that $g_\tau$ and $g^{\sss S,P}_{\sss H_j{\overline\tau}\tau}$ are
real). As one might expect, the CP-conjugate amplitude is obtained
from the original amplitude by complex-conjugating the weak phases and
leaving the strong phases unchanged (the strong phases appear in the
propagator matrix)\footnote{Actually the situation is somewhat subtle.
We have derived the expression for $\overline{\cal A}$ by determining
the CP-conjugates of the initial and final states and then calculating
the corresponding amplitude. Alternatively, one could reverse the
signs on all weak phases $\phi_i$ ($\phi_{\sss A_t}$, $\phi_{\sss A_b}$, etc.)
and recalculate ${\cal A}$. Under the transformation $\phi_i\to
-\phi_i$, one finds that $i{\rm Im}\widehat{\Pi}_{13}\to -i{\rm
Im}\widehat{\Pi}_{13}$ and $i{\rm Im}\widehat{\Pi}_{23}\to -i{\rm
Im}\widehat{\Pi}_{23}$; i.e., some of the strong phases change sign.
As a result, the 1-3 and 2-3 elements of the Higgs propagator also
change sign. A careful consideration of sign changes that occur
simultaneously in some of the coupling constants
($g_{\sss H_3\tilde{t}_2^*\tilde{t}_1}\to
-g^*_{\sss H_3\tilde{t}_2^*\tilde{t}_1}$, $g^S_{\sss H_3\overline{\tau}\tau}\to
-g^S_{\sss H_3\overline{\tau}\tau}$, $g^P_{\sss H_1\overline{\tau}\tau}\to
-g^P_{\sss H_1\overline{\tau}\tau}$ and $g^P_{\sss H_2\overline{\tau}\tau}\to
-g^P_{\sss H_2\overline{\tau}\tau}$) shows that the expressions given for
$\overline{B}$ and $\overline{C}$ are precisely correct.}. The
expression for $\left|\overline{\cal A}\right|^2$ is obtained from
Eq.~(\ref{eq:ampsquared}) by the replacement
$(B,C)\to(\overline{B},\overline{C})$. (Note that the explicit `$i$'
appearing in Eq.~(\ref{eq:ampsquared}) does {\em not} change sign
under CP conjugation.)

\section{CP Asymmetries}

\label{sec:asymmetries}

We can form three CP-violating asymmetries from the expressions for
$\left|{\cal A}\right|^2$ and $\left|\overline{\cal A}\right|^2$,
using various combinations of $\tau^\pm$ polarizations to extract
different terms. The first asymmetry is the usual rate asymmetry, for
which it is assumed that spins are not measured. The second asymmetry
is a ``single-spin'' asymmetry. It requires the measurement of a
single $\tau$ spin in the decay and is sensitive to the real part of
$BC^*$. The third asymmetry is a triple-product asymmetry and
requires the measurement of both spins. It is sensitive to the
imaginary part of $BC^*$. 

In general, the construction of the CP asymmetries proceeds in two
steps. In the first step, we sum over the $\tau^\pm$ spins (possibly
in an asymmetric manner) for the process $\tilde{t}_2^-\to
\tilde{t}_1^- \tau^-\tau^+$. We denote the resulting expression for
the amplitude squared by ``$\left.\left|{\cal
A}\right|^2\right|_{\mbox{\scriptsize (a)}}$,'' where `(a)' refers to
the specific prescription employed when summing over the spins. In the
second step we construct $\left.\left|\overline{\cal
A}\right|^2\right|_{\mbox{\scriptsize (a)}}$, the analogous quantity
for the CP-conjugate of the process, and subtract it from
$\left.\left|{\cal A}\right|^2\right|_{\mbox{\scriptsize (a)}}$. The
resulting expression is odd under CP. We also integrate over the
invariant mass $M$ of the $\tau^\pm$ pair, starting at some point
above the $Z$ resonance but including one or more Higgs resonances.
The CP asymmetries can in principle be increased by a judicious choice
for the range of this integration. Assuming $M>M_{\mbox{\scriptsize
min}}$, where $M_{\mbox{\scriptsize min}}\geq 2m_\tau$,\footnote{In
practice, we always assume that $M_{\mbox{\scriptsize min}}$ is above
the $Z$ resonance so that we can ignore $Z$ contributions.} we
generically define the width associated with the a$^{\mbox{\scriptsize
th}}$ prescription for summing over spins as
\begin{eqnarray}
      \Gamma_{\sss M_{\mbox{\scriptsize min}}}^{\mbox{\scriptsize (a)}} & 
          \equiv & \frac{1}{128 \pi^3 m_{\tilde{t}_2}^3} 
           \int_{\sss M_{\mbox{\scriptsize min}}}^{m_{\tilde{t}_2}-m_{\tilde{t}_1}}
	   \frac{dM}{M}\left(\left.\left|
		{\cal A}\right|^2\right|_{\mbox{\scriptsize (a)}}\right) \nn \\
	     & & \hskip0.4truein \times \left[\lambda\left(m_{\tilde{t}_2}^2,m_{\tilde{t}_1}^2,M^2\right)
	     \lambda\left(M^2,m_\tau^2,m_\tau^2\right)\right]^{1/2} .
	\label{eq:Gamma_generic}
\end{eqnarray}
The width for the CP-conjugate process is obtained by the replacement 
$(B,C)\to(\overline{B},\overline{C})$ and is denoted 
$\overline{\Gamma}_{\sss M_{\mbox{\scriptsize min}}}^{\mbox{\scriptsize (a)}}$. 


\subsection{Rate Asymmetry}

The first CP asymmetry is the usual rate asymmetry and is obtained by
summing symmetrically over the spin states $s_1$ and $s_2$ to yield
\begin{eqnarray}
      \left.\left|{\cal A}\right|^2\right|_{\mbox{\scriptsize rate}}
       \equiv \sum_{\mbox{\scriptsize{spins}}} \left|{\cal A}(s_1,s_2)\right|^2
       = 2\left|B\right|^2\left(M^2-4m_\tau^2\right)+
	     2\left|C\right|^2 M^2 \; .
	  \label{eq:Asq_rate}
\end{eqnarray}
This expression may be inserted in Eq.~(\ref{eq:Gamma_generic}) to
obtain $\Gamma_{\sss M_{\mbox{\scriptsize min}}}$, and similarly for
$\overline{\Gamma}_{\sss M_{\mbox{\scriptsize min}}}$ (we drop the `rate'
superscripts in this case). The resulting CP asymmetry is then
\begin{eqnarray}
      A_{\mbox{\scriptsize{CP}}}^{\mbox{\scriptsize rate}} \equiv 
           \frac{\Gamma_{\sss M_{\mbox{\scriptsize min}}}^{\mbox{\scriptsize}}
	     -\overline{\Gamma}_{\sss M_{\mbox{\scriptsize min}}}^{\mbox{\scriptsize}}}
	    {\Gamma_{\sss M_{\mbox{\scriptsize min}}}^{\mbox{\scriptsize}}
	     +\overline{\Gamma}_{\sss M_{\mbox{\scriptsize min}}}^{\mbox{\scriptsize}}} \; .
	  \label{eq:CP_rate}
\end{eqnarray}

A non-zero value for this asymmetry requires both a weak-phase
difference and a strong-phase difference between the diagrams
contributing to the decay. Since we obtain our strong phases from
absorptive parts of the Higgs boson propagator, simultaneously
non-negligible values for the asymmetry and the rate itself require
the interference of two or three nearby Higgs boson resonances.

We have found numerically that the rate asymmetry tends to be very
small for $\tilde{t}_2\to \tilde{t}_1 \tau^-\tau^+$, with values
typically at the sub-one percent level. There are a few factors that
conspire to make the rate asymmetries small; these are described in
the Appendix. At this point we simply note that this particular
observable will not be a very useful tool for the study of SUSY CP
violation in this decay process. It turns out, however, that the
measurement of the spin of one or both final-state leptons can lead to
a dramatic increase in the sensitivity of $\tilde{t}_2\to \tilde{t}_1
\tau^-\tau^+$ to CP phases. The two asymmetries described below use
lepton spins to probe the cross-terms containing $B$ and $C$ in
Eq.~(\ref{eq:Amp_sq}).\footnote{We present our results as asymmetries
depending on the spins of the $\tau^-$ and $\tau^+$.  In practice,
experimentalists would use the decay products of the taus to probe
their spins.  In that case, one could construct CP asymmetries
depending on the momenta of the decay products.  See Ref.~\cite{wkn}
for an explicit comparison of `theoretical' triple-product asymmetries
involving the $D^*$ polarization states in $B\to
D^*\ell\overline{\nu}$ with more `experimental' asymmetries that use
only the momenta in the four-body final state of the decay $B\to
(D^*\to D\pi) \ell\overline{\nu}$.}  These asymmetries can be quite
large numerically.

\subsection{Single-spin Asymmetry}

We first construct a CP asymmetry that uses the polarization of only
one of the leptons. This single-spin asymmetry extracts the piece in
$\left|{\cal A}\right|^2$ that is proportional to the real part of
$BC^*$. In constructing this asymmetry, it is useful to calculate
$\left|{\cal A}(s_1,s_2)\right|^2$ explicitly for the different spin
combinations of the $\tau^\pm$. The spin four-vectors $s_1$ and $s_2$
are determined in the usual manner by Lorentz-transforming unit
three-vectors from the rest frames of the particle and antiparticle.
Let the unit three-vector $\hat{s}$ denote a specific spin state in
the rest frame of the $\tau^+$ or $\tau^-$. Lorentz-transforming
$\hat{s}$ to a reference frame in which the $\tau^+$ or $\tau^-$ has
four-momentum $p^\mu=(E,\vec{p})$ yields
\begin{eqnarray}
      s^\mu = \left(\frac{\vec{p}\cdot\hat{s}}{m},
              \hat{s}+\frac{\left(\vec{p}\cdot\hat{s}\right)\vec{p}}{m(E+m)}\right) \; .
\end{eqnarray}
For a given set of momenta in the process $\tilde{t}_2\to \tilde{t}_1
\tau^-\tau^+$ there are four spin combinations that one could
consider, corresponding to the four combinations of $\pm\hat{s}_1$ and
$\pm\hat{s}_2$. In constructing the single-spin asymmetry it is
convenient to work in the helicity basis, taking the unit
three-vectors to be aligned or anti-aligned with the lepton
three-momenta in the rest frame of the lepton pair. This choice also
maximizes the single-spin asymmetry. (In the following subsection we
will construct triple products using the spins $s_1$ and $s_2$. In
that case it will be convenient to employ a basis in which the unit
three-vectors are perpendicular to the lepton three-momenta.)

Since $\left|{\cal A}\right|^2$ is Lorentz-invariant, we may choose
any convenient frame of reference. We work in the rest frame of the
$\tau^\pm$ pair, so that $p_{1,2}^{*\mu}=(E,\pm \vec{p})$, with
$\vec{p}$ denoting the three-momentum of the $\tau^-$ in that
frame\footnote{In the remainder of this work we will refer to the rest
frame of the $\tau^\pm$ pair as the rest frame of the decaying
(virtual) Higgs.}. Defining $\hat{s}_1=\vec{p}/|\vec{p}|$ and
Lorentz-transforming $\pm\hat{s}_1$ from the $\tau^-$ rest frame to
the Higgs rest frame, we have
$s_1^{*\pm\mu}=\pm\left(\left|\vec{p}\right|/m_\tau,
E\vec{p}/(m_\tau|\vec{p}|)\right)$. Similarly, for $\tau^+$ spins
that are in the $\pm\vec{p}/|\vec{p}|$ direction in the $\tau^+$ rest
frame, we have $s_2^{*\pm\mu}=\pm\left(-\left|\vec{p}\right|/m_\tau,
E\vec{p}/(m_\tau|\vec{p}|)\right)$ in the Higgs rest frame. Explicit
calculations for the four spin combinations yield
\begin{eqnarray}
      \left|{\cal A}(\pm,\pm)\right|^2 & = & 0 \; ,\label{eq:appmm}\\
      \left|{\cal A}(\pm,\mp)\right|^2 & = & \left|B\right|^2\left(M^2-4m_\tau^2\right)+
	     \left|C\right|^2 M^2 \nn \\
	     & & \mp\left(BC^*+B^*C\right)M\sqrt{M^2-4m_\tau^2} \; .
	\label{eq:apmmp}
\end{eqnarray}
Summing all four spin combinations reproduces the result 
in Eq.~(\ref{eq:Asq_rate}), as expected. 

Taking the difference of the two non-zero spin combinations yields
\begin{eqnarray}
      \left.\left|{\cal A}\right|^2\right|_{\mbox{\scriptsize spin}}
       \equiv \left|{\cal A}(+,-)\right|^2 - \left|{\cal A}(-,+)\right|^2
       = -4\;{\rm Re}\left(BC^*\right)M\sqrt{M^2-4m_\tau^2} \; .
	  \label{eq:Asq_spin}
\end{eqnarray}
The resulting CP asymmetry is given by
\begin{eqnarray}
      A_{\mbox{\scriptsize{CP}}}^{\mbox{\scriptsize{spin}}} \equiv 
           \frac{\Gamma_{\sss M_{\mbox{\scriptsize min}}}^{\mbox{\scriptsize spin}}
	     -\overline{\Gamma}_{\sss M_{\mbox{\scriptsize min}}}^{\mbox{\scriptsize spin}}}
	    {\Gamma_{\sss M_{\mbox{\scriptsize min}}}^{\mbox{\scriptsize}}
	     +\overline{\Gamma}_{\sss M_{\mbox{\scriptsize min}}}^{\mbox{\scriptsize}}} \; ,
	   \label{eq:acpspin}
\end{eqnarray}
in which we have normalized the asymmetry using the total widths
$\Gamma_{\sss M_{\mbox{\scriptsize min}}}^{\mbox{\scriptsize}}$ and
$\overline{\Gamma}_{\sss M_{\mbox{\scriptsize min}}}^{\mbox{\scriptsize}}$.
Note that this asymmetry only requires the measurement of a single
spin, since the $++$ and $--$ combinations are zero. As was the case
for the rate asymmetry, this single-spin asymmetry requires the
interference of two amplitudes with a non-zero relative strong phase.

In contrast to the rate asymmetry, the single-spin asymmetry can be
fairly large numerically since it does not suffer from the same
cancellations as those of the rate asymmetry.

\subsection{Triple-product Asymmetry}

\begin{figure}[t]
\begin{center}
\resizebox{2.5in}{!}{\includegraphics*{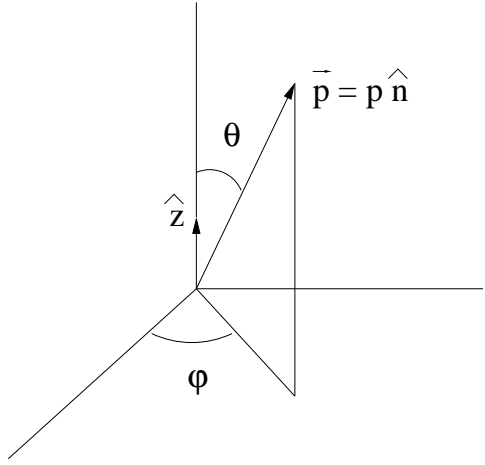}}
\caption{Definitions of angular variables and unit vectors used in the
calculation of the triple-product CP asymmetry. The unit vector
$\hat{z}$ specifies the direction of the virtual Higgs as seen in the
$\tilde{t}_2$ rest frame. The vector $\vec{p}=p\, \hat{n}$ specifies the $\tau^-$
momentum in the Higgs rest frame.}
\label{fig:triple_defs}
\end{center}
\end{figure}

For the triple-product CP asymmetry we must use a different basis for
the spins $s_{1,2}$. We work in the rest frame of the virtual Higgs
(i.e., in the rest frame of the $\tau^\pm$ pair), defining the $+z$
direction in that frame to be the direction of the Higgs as seen in
the $\tilde{t}_2$ rest frame. We let the unit vector $\hat{z}$ denote
this direction (see Fig.~\ref{fig:triple_defs}). We let the unit
vector $\hat{n}$ denote the direction of the $\tau^-$ in the Higgs
rest frame. To specify the spin four-vectors of the $\tau^\pm$, we
start by defining the corresponding unit three-vectors in the
individual rest frames of the $\tau^-$ and the $\tau^+$. These are
then Lorentz-transformed to the Higgs rest frame. In the individual
$\tau^-$ and $\tau^+$ rest frames we define the `$+$' and `$-$' spin
states as
\begin{eqnarray}
      \hat{s}_1^\pm & = & \pm \frac{\hat{z}-\hat{n}\cos\theta}
	  {\left|\hat{z}-\hat{n}\cos\theta\right|}
	  =\pm\frac{\hat{z}-\hat{n}\cos\theta}
	  {\sin\theta}\; , \label{eq:spin1}\\
      \hat{s}_2^\pm & = & \pm\frac{\hat{z}\times\hat{n}}
	  {\left|\hat{z}\times\hat{n}\right|} = 
	    \pm\frac{\hat{z}\times\hat{n}}{\sin\theta} \; , \label{eq:spin2}
\end{eqnarray}
where $\theta$ is defined in Fig.~\ref{fig:triple_defs}. Then the
vectors $\hat{s}_1^\pm$, $\hat{s}_2^\pm$ and $\hat{n}$ are mutually
perpendicular. The `$\epsilon$' term may then be extracted from
Eq.~(\ref{eq:ampsquared}) by the following prescription:
\begin{eqnarray}
      \left.\left|{\cal A}\right|^2\right|_{\mbox{\scriptsize TP}} & \equiv &
        - \frac{1}{4\pi} \int \left[\left|{\cal A}(+,+)\right|^2
	  +\left|{\cal A}(-,-)\right|^2
	  -\left|{\cal A}(+,-)\right|^2
	  -\left|{\cal A}(-,+)\right|^2\right] d\Omega \nn \\
	& = & 4\;{\rm Im}\left(BC^*\right)M\sqrt{M^2-4m_\tau^2} \; ,
	   \label{eq:asqtriple}
\end{eqnarray}
where the angular integration is performed in the Higgs rest frame,
and where the $\pm$ spins are defined as in Eqs.~(\ref{eq:spin1}) and
(\ref{eq:spin2}). (This basis is different than that employed in
Eqs.~(\ref{eq:appmm}) and (\ref{eq:apmmp}).)  Inserting
Eq.~(\ref{eq:asqtriple}) in Eq.~(\ref{eq:Gamma_generic}), and
similarly for the CP-conjugate process, we form the CP-odd triple
product asymmetry as follows:
\begin{eqnarray}
      A_{\mbox{\scriptsize{CP}}}^{\mbox{\scriptsize{TP}}} \equiv 
           \frac{\Gamma_{\sss M_{\mbox{\scriptsize min}}}^{\mbox{\scriptsize TP}}
	     -\overline{\Gamma}_{\sss M_{\mbox{\scriptsize min}}}^{\mbox{\scriptsize TP}}}
	    {\Gamma_{\sss M_{\mbox{\scriptsize min}}}^{\mbox{\scriptsize}}
	     +\overline{\Gamma}_{\sss M_{\mbox{\scriptsize min}}}^{\mbox{\scriptsize}}} \; .
	   \label{eq:acptriple}
\end{eqnarray}
Note that the prescription used to define 
$\left.\left|{\cal A}\right|^2\right|_{\mbox{\scriptsize TP}}$ is equivalent,
in this spin basis, to the definition 
\begin{eqnarray}
      \Gamma_{\sss M_{\mbox{\scriptsize min}}}^{\mbox{\scriptsize TP}} =
        \Gamma_{\sss M_{\mbox{\scriptsize min}}}\left(\epsilon(p_1,s_1,p_2,s_2)>0\right)
	-\Gamma_{\sss M_{\mbox{\scriptsize min}}}\left(\epsilon(p_1,s_1,p_2,s_2)<0\right)\; .
\end{eqnarray}

The triple-product asymmetry does not require a relative strong phase between
interfering amplitudes. In fact, the triple-product asymmetry does not
even necessarily require the interference of two different Higgs resonances --
one resonance suffices as long as there is scalar-pseudoscalar mixing among 
the Higgs bosons.

\section{Results}

\label{sec:results}

As has been noted above and discussed in the Appendix, the rate
asymmetry in $\tilde{t}_2\to \tilde{t}_1 \tau^-\tau^+$ will in general
be very small, so we do not consider it further here. The single-spin
CP asymmetry can be large, but requires the interference of two or
three nearby Higgs resonances. If the resonances are spread far apart,
the product of any two interfering amplitudes will be suppressed,
since one amplitude will be much smaller than the other. The
strong-phase difference in this CP asymmetry is due to the width
difference between the contributing Higgs bosons as well as the
relative values of their masses (see Eq.~(\ref{eq:str_phase})).  The
triple-product CP asymmetry does not require a strong-phase difference
and can receive large contributions from single resonances (due to
scalar-pseudoscalar mixing) and from the interference between two
resonances. The Appendix contains further discussions of these two
distinct types of contributions.

In SUSY it is somewhat natural to have nearby Higgs resonances, so
that it is not unreasonable to hope for large single-spin and
triple-product asymmetries. At tree level there is a well-known
relation between the masses of the heavier scalar Higgs ($H$) and the
pseudoscalar Higgs ($A$),
\begin{eqnarray}
      m_H^2-m_A^2 \simeq m_Z^2\sin^22\beta \; ,
       \label{eq:tree_level_masses}
\end{eqnarray}
a relation which is valid in the limit $m_A^2\gg m_Z^2$
\cite{higgs_tree}. For large $\tan\beta$, $\sin^22\beta\approx 0$ and
thus $m_H^2\approx m_A^2$. If CP is not conserved, loop corrections
lead to the mixing of the scalar and pseudoscalar Higgs bosons
\cite{carena2002}. This mixing can lead to the breaking of the near
degeneracy in masses, as has been explored in some detail in
Ref.~\cite{bernabeu}. Nevertheless, for some combinations of
parameters (in particular, it appears, if $|\mu|$ is not too large)
there remains a near degeneracy of the two heavier Higgs bosons.

In the following we choose a few points in the SUSY parameter space
and use these to illustrate some of the features of the single-spin
and triple-product asymmetries. Our treatment is certainly not
exhaustive, but is meant simply to point to the usefulness of using
these asymmetries as a tool for exploring SUSY CP violation. We
choose values for the charged Higgs mass and $\tan\beta$ that are
consistent with the recent bound from Belle, $\tan\beta/m_{\sss H^\pm}\lsim
0.146~\mbox{GeV}^{-1}$ \cite{belle}.

The asymmetries
$A_{\mbox{\scriptsize{CP}}}^{\mbox{\scriptsize{spin}}}$ and
$A_{\mbox{\scriptsize{CP}}}^{\mbox{\scriptsize{TP}}}$ defined in
Eqs.~(\ref{eq:acpspin}) and (\ref{eq:acptriple}) have been integrated
over $M$, where $M=\sqrt{(p_1+p_2)^2}$ is the invariant mass of the
$\tau^\pm$ pair (see Eq.~(\ref{eq:Gamma_generic})). Since the
$\tau$'s are produced by the $s$-channel decays of Higgs bosons (see
Fig.~\ref{fig:feyn_diag}) , the differential widths (in $M$) for the
decays $\tilde{t}_2^\pm\to \tilde{t}_1^\pm \tau^-\tau^+$ are expected
to exhibit resonant peaks in the vicinities of the neutral Higgs
masses; i.e., when $M\approx m_{\sss H_i}$, with $i=1,2,3$. The CP
asymmetries that we have constructed have a sensitive dependence on
the resonance structure of the decay. For example, as mentioned
above, the single-spin asymmetry requires the interference of
different resonances, and will be suppressed if the resonances are too
far apart. The triple-product asymmetry can receive contributions from
single resonances, but in some cases the contributions due to
different resonances could be of opposite signs, leading to a
suppression of the (integrated) quantity
$A_{\mbox{\scriptsize{CP}}}^{\mbox{\scriptsize{TP}}}$.
Experimentalists may wish to enforce various cuts on the integration
over $M$ in order to optimize both the CP asymmetry and its
statistical significance.

To display some of the above features more clearly, let us define
several differential quantities. First, we define
$d\Gamma_{\mbox{\scriptsize sum}}/dM$ to be the sum of the total
(i.e., summed over spins) differential widths for the process and the
anti-process. That is,
\begin{eqnarray}
      \frac{d\Gamma_{\mbox{\scriptsize sum}}}{dM} & 
          \equiv & \frac{1}{128 \pi^3 m_{\tilde{t}_2}^3} 
	   \frac{1}{M}\left(\left.\left|
		{\cal A}\right|^2\right|_{\mbox{\scriptsize rate}}+
		\left.\left|
		\overline{\cal A}\right|^2\right|_{\mbox{\scriptsize rate}}\right)
		\nn \\
	     & & \hskip0.4truein \times \left[\lambda\left(m_{\tilde{t}_2}^2,m_{\tilde{t}_1}^2,M^2\right)
	     \lambda\left(M^2,m_\tau^2,m_\tau^2\right)\right]^{1/2} .
	\label{eq:dGamma_sum}
\end{eqnarray}
A plot of $d\Gamma_{\mbox{\scriptsize sum}}/dM$ as a function of $M$
will show resonance structure, with peaks appearing at the Higgs
masses. Furthermore, integrating $d\Gamma_{\mbox{\scriptsize
sum}}/dM$ over $M$ gives the sum of the widths for the process and the
anti-process,
\begin{eqnarray}
      \Gamma_{\mbox{\scriptsize sum}}\equiv
	   \int_{\sss M_{\mbox{\scriptsize min}}}^{m_{\tilde{t}_2}-m_{\tilde{t}_1}}
	     \frac{d\Gamma_{\mbox{\scriptsize sum}}}{dM}dM 
	=\Gamma_{\sss M_{\mbox{\scriptsize min}}}^{\mbox{\scriptsize}}
	     +\overline{\Gamma}_{\sss M_{\mbox{\scriptsize min}}}^{\mbox{\scriptsize}}\; .
\end{eqnarray}
$\Gamma_{\mbox{\scriptsize sum}}$ is the quantity that appears in the
denominator of the CP asymmetries defined in Eqs.~(\ref{eq:acpspin})
and (\ref{eq:acptriple}). Recall as well that it has an implicit
dependence on the lower limit, $M_{\mbox{\scriptsize min}}$, of the
integation in the above expression. We also define two CP-asymmetric
differential widths,
\begin{eqnarray}
      \frac{d\Gamma_{\mbox{\scriptsize CP}}^{\mbox{\scriptsize (a)}}}{dM} & 
          \equiv & \frac{1}{128 \pi^3 m_{\tilde{t}_2}^3} 
	   \frac{1}{M}\left(\left.\left|
		{\cal A}\right|^2\right|_{\mbox{\scriptsize (a)}}-
		\left.\left|
		\overline{\cal A}\right|^2\right|_{\mbox{\scriptsize (a)}}\right)
		\nn \\
	     & & \hskip0.4truein \times \left[\lambda\left(m_{\tilde{t}_2}^2,m_{\tilde{t}_1}^2,M^2\right)
	     \lambda\left(M^2,m_\tau^2,m_\tau^2\right)\right]^{1/2} ,
	\label{eq:dGamma_generic}
\end{eqnarray}
where `a' corresponds to `spin' or `TP'. When integrated over $M$,
these differential widths form the numerators in the expressions for
the CP asymmetries in Eqs.~(\ref{eq:acpspin}) and
(\ref{eq:acptriple}), so that
\begin{eqnarray}
      A_{\mbox{\scriptsize{CP}}}^{\mbox{\scriptsize{(a)}}} =
           \frac{1}{\Gamma_{\mbox{\scriptsize sum}}}
           \int_{\sss M_{\mbox{\scriptsize min}}}^{m_{\tilde{t}_2}-m_{\tilde{t}_1}}
	     \frac{d\Gamma_{\mbox{\scriptsize CP}}^{\mbox{\scriptsize (a)}}}{dM}dM \; .
	 \label{eq:int_exp_ACP}
\end{eqnarray}
In the following we will show plots of $d\Gamma_{\mbox{\scriptsize
CP}}^{\mbox{\scriptsize spin}}/dM$, $d\Gamma_{\mbox{\scriptsize
CP}}^{\mbox{\scriptsize TP}}/dM$ and $d\Gamma_{\mbox{\scriptsize
sum}}/dM$. The relation between these plots and the integrated CP
asymmetries $A_{\mbox{\scriptsize{CP}}}^{\mbox{\scriptsize{spin}}}$
and $A_{\mbox{\scriptsize{CP}}}^{\mbox{\scriptsize{TP}}}$ follows from
Eq.~(\ref{eq:int_exp_ACP}): The spin (TP) asymmetry is the ratio of
the area under the curve $d\Gamma_{\mbox{\scriptsize
CP}}^{\mbox{\scriptsize spin(TP)}}/dM$ to the area under the curve
$d\Gamma_{\mbox{\scriptsize sum}}/dM$.

\begin{figure}[t]
\begin{center}
\resizebox{5.5in}{!}{\includegraphics*{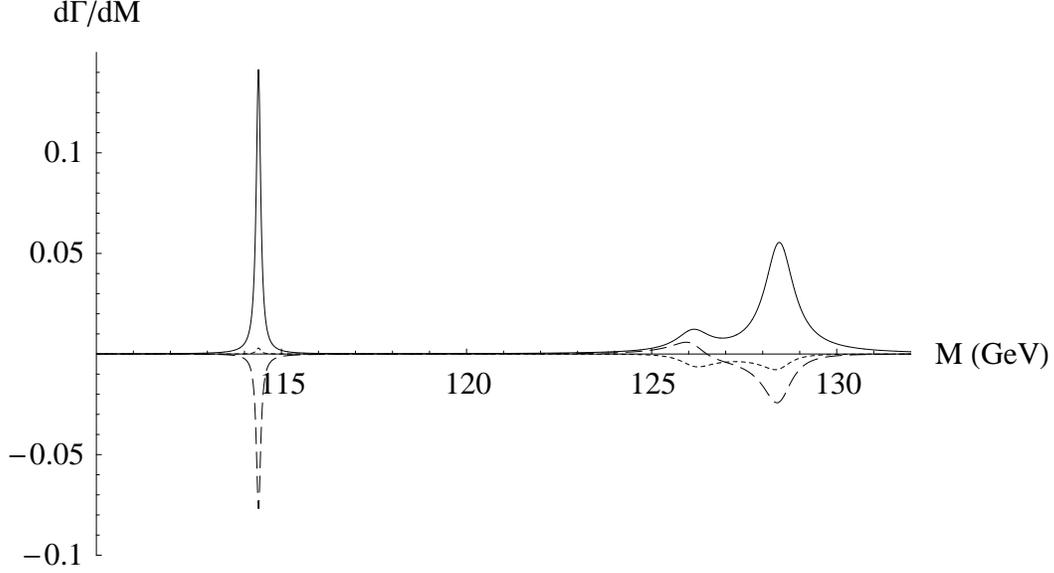}}
\caption{Plot of $\frac{d\Gamma_{\mbox{\scriptsize sum}}}{dM}$ (solid
line), $\frac{d\Gamma_{\mbox{\scriptsize CP}}^{\mbox{\scriptsize
TP}}}{dM}$ (long dashed line) and $\frac{d\Gamma_{\mbox{\scriptsize
CP}}^{\mbox{\scriptsize spin}}}{dM}$ (short dashed line) for
a case in which the two heaviest Higgs resonances (near $126.1$~GeV
and $128.5$~GeV) have a significant overlap.}
\label{fig:cl_res}
\end{center}
\end{figure}

Figure~\ref{fig:cl_res} shows plots of the differential widths
$d\Gamma_{\mbox{\scriptsize CP}}^{\mbox{\scriptsize spin}}/dM$,
$d\Gamma_{\mbox{\scriptsize CP}}^{\mbox{\scriptsize TP}}/dM$ and
$d\Gamma_{\mbox{\scriptsize sum}}/dM$ for a particular set of input
parameters. To generate these plots we have used {\tt CPsuperH} to
determine the masses and couplings of the Higgs bosons as well as the
masses of the stops. Some of the input parameters are as follows:
$m_{\sss H^\pm}=150$~GeV, $\mu=+200$~GeV, $\left|A_t\right|=600$~GeV,
$\phi_{\sss A_t}=90^\circ$, $\tan\beta=20$, $m_{\tilde{Q}_3}=350$~GeV and
$m_{\tilde{U}_3}=400$~GeV, where we have followed the notation of {\tt
CPsuperH}. (The decay process strongly depends on all of these
parameters.  Other parameters have been chosen such that the neutral
Higgs bosons cannot decay into SUSY particles.)  The only non-zero
phase is that of the trilinear coupling $A_t$. The masses of the stops
are determined by {\tt CPsuperH} to be $m_{\tilde{t}_{1}}\simeq
259$~GeV and $m_{\tilde{t}_{2}}\simeq 511$~GeV.  Similarly, the neutral
Higgs masses (widths) are $m_{\sss H_1}\simeq 114.38~(0.15)$~GeV,
$m_{\sss H_2}\simeq 126.12~(1.09)$~GeV and $m_{\sss H_3}\simeq
128.45~(0.95)$~GeV.

As noted above, to obtain the integrated CP asymmetries, we determine
the areas under the curves in Fig.~\ref{fig:cl_res} and calculate the
appropriate ratios. A comparison of the area under the dashed curve
($d\Gamma_{\mbox{\scriptsize CP}}^{\mbox{\scriptsize TP}}/dM$) to that
under the solid curve ($d\Gamma_{\mbox{\scriptsize sum}}/dM$)
indicates that the CP asymmetry will be relatively large in this case. A
similar analysis for the single-spin asymmetry leads one to the
conclusion that that asymmetry will be somewhat smaller. Our numerical
results confirm these qualitative observations. Integrating over
all three Higgs resonances we have
$A_{\mbox{\scriptsize{CP}}}^{\mbox{\scriptsize{spin}}}\simeq -14\%$ and
$A_{\mbox{\scriptsize{CP}}}^{\mbox{\scriptsize{TP}}}\simeq -31\%$.

Consider also a few other technical details associated with this
example:
\begin{enumerate}

\item One might wonder why the resonance evident in
$d\Gamma_{\mbox{\scriptsize sum}}/dM$ near the heaviest Higgs at
$128$~GeV is larger than that near $126$~GeV. It turns out that the
difference in the strengths of these resonances is due primarily to
differences in the Higgs-stop-stop couplings.

\item In this example the two heaviest Higgs resonances have a
relatively significant overlap, allowing for some amount of
interference between the resonances. The single-spin asymmetry
requires such interference, while the triple-product asymmetry can
also receive contributions from single resonances.

\item Including only the heavier two resonances in the integration
over $M$ increases
$A_{\mbox{\scriptsize{CP}}}^{\mbox{\scriptsize{spin}}}$ (in terms of
its magnitude) to about $-19\%$, but decreases
$A_{\mbox{\scriptsize{CP}}}^{\mbox{\scriptsize{TP}}}$ (again in terms
of its magnitude) to about $-24\%$. The increase in the single-spin
asymmetry is not difficult to understand: The numerator in
Eq.~(\ref{eq:int_exp_ACP}) is nearly unchanged (because the $H_1$
resonance is so far separated from the other two and does interfere
very much), but the denominator decreases from
$\Gamma_{\mbox{\scriptsize sum}}\simeq 0.13$~GeV to
$\Gamma_{\mbox{\scriptsize sum}}\simeq 0.098$~GeV.

\item Observation of a triple-product asymmetry in the vicinity of a
single, widely-separated resonance (such as $H_1$ in this case) would
be evidence of scalar-pseudoscalar mixing in the Higgs sector.

\end{enumerate}

\begin{figure}[tbh]
\begin{center}
\resizebox{5.5in}{!}{\includegraphics*{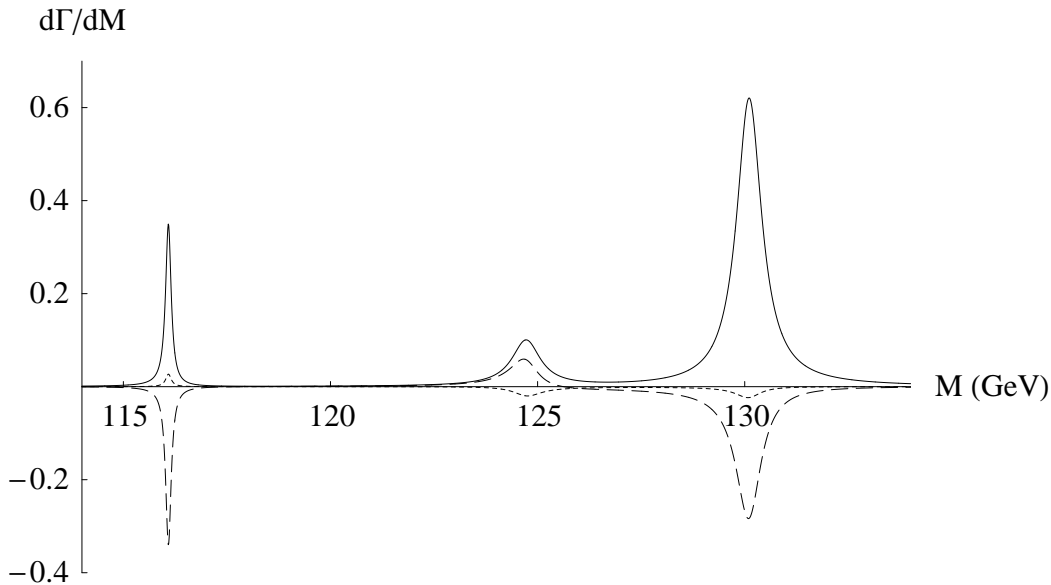}}
\caption{Plot of $\frac{d\Gamma_{\mbox{\scriptsize sum}}}{dM}$ (solid
line), $\frac{d\Gamma_{\mbox{\scriptsize CP}}^{\mbox{\scriptsize
TP}}}{dM}$ (long dashed line) and $\frac{d\Gamma_{\mbox{\scriptsize
CP}}^{\mbox{\scriptsize spin}}}{dM}$ (short dashed line) for
a case in which the two heaviest Higgs resonances are somewhat spread
apart. The input parameters for this plot are noted in the text. }
\label{fig:far_res}
\end{center}
\end{figure}

A second example is shown in Fig.~\ref{fig:far_res}. The input
parameters for this example are the same as those of the previous
example, except that $\mu$ and $A_t$ have been increased in magnitude
($\mu=+500$~GeV and $\left|A_t\right|=700$~GeV in this case). As a
result, $m_{\sss H_2}$ and $m_{\sss H_3}$ have moved farther away from each
other and $m_{\sss H_1}$ has increased.  (The neutral Higgs masses
(widths), as determined by {\tt CPsuperH}, are $m_{\sss H_1}\simeq
116.09~(0.16)$~GeV, $m_{\sss H_2}\simeq 124.73~(0.83)$~GeV and
$m_{\sss H_3}\simeq 130.11~(0.74)$~GeV, in this case.  The stop masses are
$m_{\tilde{t}_{1}}\simeq 228$~GeV and $m_{\tilde{t}_{2}}\simeq 526$~GeV.)
The CP asymmetries and integrated width in this example, when
integrated over all three (only the heavier two) resonances, are
$A_{\mbox{\scriptsize{CP}}}^{\mbox{\scriptsize{spin}}}\simeq -5\%
~(-6\%)$, $A_{\mbox{\scriptsize{CP}}}^{\mbox{\scriptsize{TP}}}\simeq
-36\% ~(-30\%)$ and $\Gamma_{\mbox{\scriptsize sum}}\simeq
0.93~(0.83)$~GeV. The single-spin asymmetry has been reduced somewhat,
compared to the previous example, due in part to the fact that the
resonances are more widely separated in this case. Also note that the
triple-product asymmetry suffers from some amount of cancellation
between the second and third resonances, since the area under the
curve near these two resonances is of opposite sign. In such a
scenario it might be experimentally preferable when measuring the
triple-product asymmetry to restrict the integration over $M$ to
smaller regions or to weight the region near the middle resonance with
a relative negative sign (to avoid the cancellation).

\begin{figure}[tbh]
\begin{center}
\resizebox{5.5in}{!}{\includegraphics*{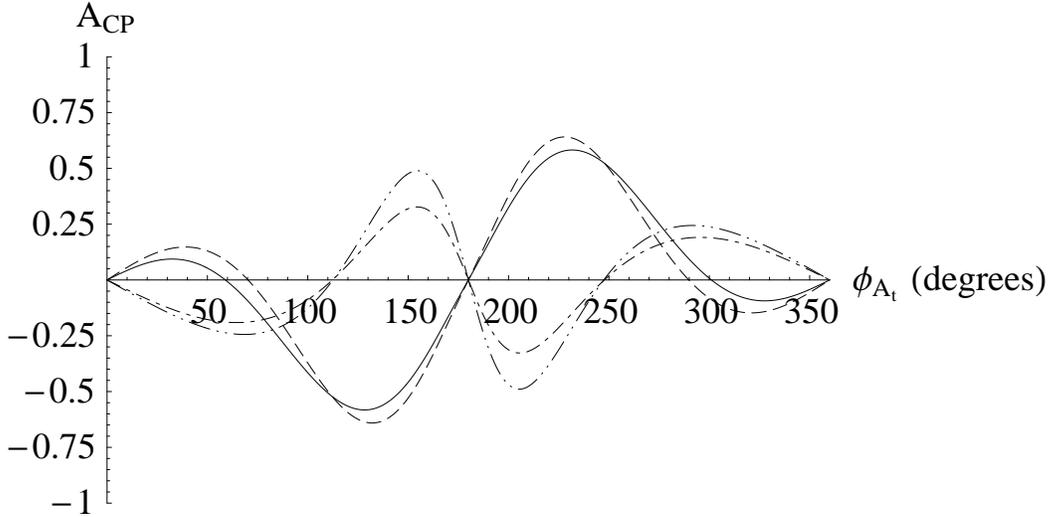}}
\caption{Plot of the integrated triple-product and single-spin
asymmetries, $A_{\mbox{\scriptsize{CP}}}^{\mbox{\scriptsize{TP}}}$ and
$A_{\mbox{\scriptsize{CP}}}^{\mbox{\scriptsize{spin}}}$, respectively,
as functions of $\phi_{\sss A_t}$. The input parameters are the same as
those used for the plot in Fig.~\ref{fig:cl_res}, except that
$\phi_{\sss A_t}$ is allowed to vary. The solid (dashed) curve corresponds
to $A_{\mbox{\scriptsize{CP}}}^{\mbox{\scriptsize{TP}}}$ integrated
over all three (only the heavier two) resonances. Similarly, the
long-short-dashed (dashed-double-dotted) curve corresponds to
$A_{\mbox{\scriptsize{CP}}}^{\mbox{\scriptsize{spin}}}$ integrated
over all three (only the heavier two) resonances.}
\label{fig:phi_at}
\end{center}
\end{figure}

Figure~\ref{fig:phi_at} shows the variation of the integrated CP
asymmetries as functions of the phase of $A_t$, $\phi_{\sss A_t}$, for the
same set of input parameters used in Fig.~\ref{fig:cl_res} (except, of
course, for $\phi_{\sss A_t}$). Note that the masses of the Higgs bosons
and stops, as well as the widths of the Higgs bosons, are themselves
functions of $\phi_{\sss A_t}$, so their values change over the course of
the plot. The four curves correspond to
$A_{\mbox{\scriptsize{CP}}}^{\mbox{\scriptsize{TP}}}$ and
$A_{\mbox{\scriptsize{CP}}}^{\mbox{\scriptsize{spin}}}$, with two
different choices for $M_{\mbox{\scriptsize min}}$, one of which
includes all three resonances and one of which includes only those
corresponding to $H_2$ and $H_3$. In this example, including the first
resonance tends to decrease the magnitude of
$A_{\mbox{\scriptsize{CP}}}^{\mbox{\scriptsize{spin}}}$, while it
sometimes increases the magnitude of
$A_{\mbox{\scriptsize{CP}}}^{\mbox{\scriptsize{TP}}}$ and sometimes
decreases it.  This is evident from the plot. Values as large as about
$64\%$ and $49\%$ are observed for the triple-product and single-spin
asymmetries, respectively.  It is possible to obtain even larger
triple-product asymmetries than those shown in Fig.~\ref{fig:phi_at}.
For example, using the same parameters as those used for
Fig.~\ref{fig:far_res}, but taking $\phi_{\sss A_t}=135^\circ$, one obtains
$A_{\mbox{\scriptsize{CP}}}^{\mbox{\scriptsize{TP}}}\simeq -92\%$ and
$\Gamma_{\mbox{\scriptsize sum}}\simeq 0.73$~GeV when integrating over
all three resonances.

We have performed a limited study of the effects of allowing other
SUSY parameters to have non-zero phases.  The phase in the trilinear
coupling $A_b$ could in principle affect our asymmetries through the
mixing of the Higgs bosons.  In practice, we have found that this
phase only seems to have a non-negligible effect when
$\left|A_b\right|$ is quite large -- of order 5-10~TeV in the cases
studied -- due to the difference in the top and bottom Yukawa
couplings.  The phases of the gaugino masses could also in principle
give contributions due to their effects on Higgs-fermion-fermion
effective vertices.  Taking the magnitudes of $M_1$, $M_2$ and $M_3$
to be approximately 200, 400 and 1000~GeV, we have found that in the
cases considered these quantities can make changes to the triple
product asymmetry that are of order $\pm 0.1$.  The effect on the
single-spin asymmetry tends to be smaller in magnitude.  

We have also considered scenarios involving different values for
$\tan\beta$ or $A_t$.  The triple-product and single-spin asymmetries
are strongly dependent on the particular values chosen for these
quantities.  In the few examples we studied, decreasing $\tan\beta$
from 20 (the value used in Figs.~\ref{fig:cl_res}-\ref{fig:phi_at})
tended to decrease the asymmetries.  We also increased $\tan\beta$ in
one case (and increased $m_{H^\pm}$ to respect the experimental bound
on $\tan\beta/m_{H^\pm}$) and found that the triple-product asymmetry
could increase in that case.  Varying the magnitude of $A_t$ (while
holding its phase constant) can also affect the asymmetries.  In the
cases studied, the asymmetries tended to be somewhat constant as
$\left| A_t\right|$ was varied between 400~GeV and $700$~GeV.  There
were more significant changes in the range $700$~GeV to 1~TeV.

We should also make a comment regarding the invariance of our results
under rephasing of the complex SUSY parameters.  As is well-known, it
is possible to rotate away two phases associated with the various
complex parameters in supersymmetry (see, for example,
Ref.~\cite{abel}).  {\tt CPsuperH} appears to use one of these degrees
of freedom to force the SUSY parameter $m_{12}^2$ to be
real~\cite{carena2002,carena2000}.  The remaining complex quantities
within {\tt CPsuperH} are the gaugino masses ($M_i$, with $i=1,2,3$),
the third-generation $A$-terms ($A_\alpha$, with $\alpha=b,t,\tau$)
and $\mu$.  Six physical CP phases may be constructed from these seven
complex parameters.  These may be taken to be
$\mbox{Arg}\left(A^*_bM_i\right)$ and $\mbox{Arg}\left(\mu
A_\alpha\right)$~\cite{abel}.  (Although our scenario is not identical
to that in Ref.~\cite{abel}, we may borrow the results from that
reference because for our calculation we may consider the first two
generations to be decoupled from the third.  The more general case is
considered in Ref.~\cite{lebedev}.)  In this work we have used the
remaining available phase rotation to set
$\mbox{Arg}\left(\mu\right)=0$, in which case the phases of the
$A$-terms are physical.  Nevertheless, an important check of our
asymmetries would be to study their behavior under a rephasing
transformation.  For example, if one wished to make $A_t$ real and
positive through the transformation $A_t\to A_te^{-i\phi_{A_t}}$, then
the phase $\phi_{A_t}$ would appear in the other complex parameters,
since one would also need to take $\mu\to \mu e^{i\phi_{A_t}}$,
$A_{b,\tau}\to A_{b,\tau}e^{-i\phi_{A_t}}$ and $M_i\to M_i
e^{-i\phi_{A_t}}$.  The asymmetries should be unchanged under such
transformations.  We have performed a numerical check of the rephasing
invariance of our asymmetries in a few particular examples and have
found them to be invariant to within the expected degree of numerical
precision.

To summarize the numerical work, we note that while the regular rate
asymmetry is expected to be very small for $\tilde{t}_2^\pm\to
\tilde{t}_1^\pm \tau^-\tau^+$, we have seen that the single-spin and
triple-product asymmetries can be quite large. In this work we have
seen examples where the former can be of order $50\%$ and the latter
of order $90\%$. Such large values indicate that these asymmetries
provide a promising avenue for the measurement of CP violation within
SUSY. In particular, the observables studied here are very sensitive
to the phase of the trilinear coupling, $A_t$, as is evidenced in
Fig.~\ref{fig:phi_at}. We have also seen that the differential CP
asymmetries can have an interesting functional dependence on $M$, the
invariant mass of the $\tau^\pm$ pair. Depending on the particular set
of circumstances, experimentalists may wish to consider certain
resonances separately, or to find other ways to optimize the CP-odd
signals without sacrificing their statistical significance.

\section{Conclusions}

\label{sec:conclusions}

Many physicists believe that supersymmetry (SUSY) will be discovered
at future high-energy colliders. If so, first measurements will
involve CP-conserving quantities such as the masses of SUSY particles,
etc. However, SUSY also contains a number of CP-violating parameters.
It is only through their measurement that one will be able to identify
the type of SUSY theory which is found in nature.

In this paper we have studied CP violation in the decay ${\tilde t}_2
\to {\tilde t}_1 \tau^- \tau^+$. We have shown that two CP asymmetries
can be quite large in some regions of the SUSY parameter space (up to
90\% for the parameters we have chosen). They are the single-spin and
triple-product asymmetries, and involve the measurement of one or both
of the $\tau$ spins. Both of these asymmetries depend almost entirely
on a single CP-violating parameter, the phase of $A_t$, $\phi_{\sss A_t}$.
Thus, the measurement of the CP asymmetries in this decay will allow
one to extract or constrain $\phi_{\sss A_t}$. (Future work will involve
other channels, sensitive to other SUSY phases.)


\begin{acknowledgments}
We thank Alakabha Datta for the question which led to this study and
acknowledge helpful discussions with K.R.S. Balaji, G. B\'elanger and
H. Logan. We also thank J.S. Lee for helpful correspondence.  This
work was financially supported by NSERC of Canada. The work of
K.K. was supported in part by the U.S.\ National Science Foundation
under Grants PHY--0301964 and PHY--0601103.
\end{acknowledgments}


\appendix*

\section{Some Approximate Expressions}

In this appendix we provide some approximate expressions
that help to clarify the discussions of the three
asymmetries.

\subsection{Rate Asymmetry}

In the text it is noted that the rate asymmetry is suppressed for
$\tilde{t}_2\to \tilde{t}_1 \tau^-\tau^+$. To see why this is the
case, consider the difference in the amplitudes-squared (which appears
in the numerator of the expression for the rate asymmetry,
Eq.~(\ref{eq:CP_rate})),
\begin{eqnarray}
      \left.\left|{\cal A}\right|^2\right|_{\mbox{\scriptsize rate}}
       - \left.\left|\overline{\cal A}\right|^2\right|_{\mbox{\scriptsize rate}}
       & = & -\frac{64\pi v^2}{\lambda^{1/2}(1,\kappa_\tau,\kappa_\tau)}
        \sum_{k>i}\left|
	g_{\sss H_i\tilde{t}_2^*\tilde{t}_1}g_{\sss H_k\tilde{t}_2^*\tilde{t}_1}\right|
	 \nn \\
       & & \hskip0.4truein \times
	\sin\left(\alpha_i-\alpha_k\right) {\rm Im}\left[
	  \left(D {\rm Im}
	    \widehat{\Pi}^{\tau\tau} D^*\right)_{ik}\right] \; .
	\label{eq:amp_diff}
\end{eqnarray}
To derive this expression we have made use of the definition of
${\rm Im}\widehat{\Pi}^{\tau\tau}_{ij}(M^2)$ in 
Eq.~(\ref{eq:ImPitautau}); we have also set 
\begin{eqnarray}
      \alpha_i\equiv\arg(g_{\sss H_i\tilde{t}_2^*\tilde{t}_1}) \; .
      \label{eq:alpha_def}
\end{eqnarray}

The above expression may be simplified by assuming that the
off-diagonal terms in the propagator are small (although not
negligible) compared to the diagonal elements.  Expanding
Eq.~(\ref{eq:amp_diff}) under this assumption, we find
\begin{eqnarray}
      \left.\left|{\cal A}\right|^2\right|_{\mbox{\scriptsize rate}}
       - \left.\left|\overline{\cal A}\right|^2\right|_{\mbox{\scriptsize rate}}
       & \simeq & -\frac{64\pi v^2}{\lambda^{1/2}(1,\kappa_\tau,\kappa_\tau)}
       \sum_{i,k}\left|
	g_{\sss H_i\tilde{t}_2^*\tilde{t}_1}g_{\sss H_k\tilde{t}_2^*\tilde{t}_1}\right|
	\sin\left(\alpha_i-\alpha_k\right) \times \frac{1}{\left|X_{ik}\right|^2}
	 \nn \\
	& & \times
	\left(M^2-m_{\sss H_k}^2\right)\left(
	  {\rm Im}\widehat{\Pi}_{ik}{\rm Im}\widehat{\Pi}^{\tau\tau}_{ii}-
	  {\rm Im}\widehat{\Pi}_{ii}{\rm Im}\widehat{\Pi}^{\tau\tau}_{ik}\right)
    \label{eq:A_A_v1} \\
       & = & -\frac{64\pi v^2}{\lambda^{1/2}(1,\kappa_\tau,\kappa_\tau)}
       \sum_{i,k}\left|
	g_{\sss H_i\tilde{t}_2^*\tilde{t}_1}g_{\sss H_k\tilde{t}_2^*\tilde{t}_1}\right|
	\sin\left(\alpha_i-\alpha_k\right) \times \frac{1}{\left|X_{ik}\right|^2}
	 \nn \\
	& & \times
	\left(M^2-m_{\sss H_k}^2\right)\left(
	  {\rm Im}\widehat{\Pi}_{ik}^{bb}{\rm Im}\widehat{\Pi}^{\tau\tau}_{ii}-
	  {\rm Im}\widehat{\Pi}_{ii}^{bb}{\rm Im}\widehat{\Pi}^{\tau\tau}_{ik}\right)\; ,
    \label{eq:A_A_v2} 
\end{eqnarray}
where
\beq
   X_{ik} \equiv \left(M^2-m_{\sss H_i}^2+i \, {\rm Im}\widehat{\Pi}_{ii}\right)
    \left(M^2-m_{\sss H_k}^2+i \, {\rm Im}\widehat{\Pi}_{kk}\right) \; .
\eeq
In writing down these expressions, we have omitted a class of terms
that could be important if all three resonances are close together.

The expression in Eq.~(\ref{eq:A_A_v2}) is small for a few
reasons. First, a cancellation has occurred so that only self-energy
loops with $b$ quarks contribute to the required strong phase
difference (the $\tau$ contributions disappear in going from
Eq.~(\ref{eq:A_A_v1}) to Eq.~(\ref{eq:A_A_v2}))\footnote{Note the
importance of keeping the off-diagonal propagator terms in this
case.}. Second, the terms 
${\rm Im}\widehat{\Pi}_{ik}^{bb}{\rm Im}\widehat{\Pi}^{\tau\tau}_{ii}$
and 
$-{\rm Im}\widehat{\Pi}_{ii}^{bb}{\rm Im}\widehat{\Pi}^{\tau\tau}_{ik}$
tend to cancel each other.  Part of the reason for this is that
the coupling constants $g^{\sss S,P}_{\sss H_i\overline{b}b}$
and $g^{\sss S,P}_{\sss H_i\overline{\tau}\tau}$ are equal at tree-level.

\subsection{Single-spin Asymmetry}

In our numerical work we keep the full $3\times 3$ Higgs propagator,
but to understand some of the dynamics for the single-spin asymmetry
it is sufficient to keep only the diagonal terms. In that case we
obtain
\begin{eqnarray}
      \left.\left|{\cal A}\right|^2\right|_{\mbox{\scriptsize spin}}
       -\left.\left|\overline{\cal A}\right|^2\right|_{\mbox{\scriptsize spin}}
       & \simeq & 8 M \sqrt{M^2-4 m_\tau^2} v^2 g_\tau^2 \nn \\
       & & \times \sum_{j\neq k}
	  \frac{\left|
	  g_{\sss H_j\tilde{t}_2^*\tilde{t}_1}g_{\sss H_k\tilde{t}_2^*\tilde{t}_1}\right|
	  g^S_{\sss H_k\overline{\tau}\tau}g^P_{\sss H_j\overline{\tau}\tau}
          \cos\left(\alpha_k-\alpha_j\right)}
	  {\left[\left(M^2-m_{\sss H_k}^2\right)^2+\Gamma_k^2 m_{\sss H_k}^2\right]
	    \left[\left(M^2-m_{\sss H_j}^2\right)^2+\Gamma_j^2 m_{\sss H_j}^2\right]}
	                \nn \\
       & & \times 
	  \left[\Gamma_k m_{\sss H_k}\left(M^2-m_{\sss H_j}^2\right)-
	    \Gamma_j m_{\sss H_j}\left(M^2-m_{\sss H_k}^2\right)\right] \; ,
\end{eqnarray}
where we have also made the approximation ${\rm
Im}\widehat{\Pi}_{jj}(M^2)\simeq \Gamma_j m_{\sss H_j}$, with
$\Gamma_j\equiv \Gamma(H_j)$. If we define the strong phases
$\delta_j(M^2)$ as follows
\begin{eqnarray}
      \cos\delta_j+i\sin\delta_j \equiv
        \frac{\left(M^2-m_{\sss H_j}^2\right)-i\Gamma_jm_{\sss H_j}}
	{\sqrt{\left(M^2-m_{\sss H_j}^2\right)^2+\Gamma_j^2 m_{\sss H_j}^2}} \; ,
	\label{eq:str_phase}
\end{eqnarray}
we have
\begin{eqnarray}
      \left.\left|{\cal A}\right|^2\right|_{\mbox{\scriptsize spin}}
       -\left.\left|\overline{\cal A}\right|^2\right|_{\mbox{\scriptsize spin}}
       & \simeq & -8 M \sqrt{M^2-4 m_\tau^2} v^2 g_\tau^2 \nn \\
       & & \!\!\!\!\!\!\!\!\!\!\!\!\!\!\!\!\!\!\!\!\!\!\!\! \times \sum_{j\neq k}
	  \frac{\left|
	  g_{\sss H_j\tilde{t}_2^*\tilde{t}_1}g_{\sss H_k\tilde{t}_2^*\tilde{t}_1}\right|
	  g^S_{\sss H_k\overline{\tau}\tau}g^P_{\sss H_j\overline{\tau}\tau}
          \cos\left(\alpha_k-\alpha_j\right)\sin\left(\delta_k-\delta_j\right)}
	  {\sqrt{\left[\left(M^2-m_{\sss H_k}^2\right)^2+\Gamma_k^2 m_{\sss H_k}^2\right]
	    \left[\left(M^2-m_{\sss H_j}^2\right)^2+\Gamma_j^2 m_{\sss H_j}^2\right]}} .
	  \label{eq:Asinglespin_approx}
\end{eqnarray}
In this form we can see clearly that the single-spin asymmetry depends
on the sine of the relative strong phase between the interfering
resonances. But it is perhaps somewhat counterintuitive that the
single-spin asymmetry depends on the {\em cosine} of
$\alpha_k-\alpha_j$, the weak phase difference.

To understand the presence of `$\cos(\alpha_k-\alpha_j)$' in
Eq.~(\ref{eq:Asinglespin_approx}), consider first the CP-invariant
limit. In this case there is no scalar-pseudoscalar mixing -- two of
the Higgs bosons are pure scalars and one is a pure pseudoscalar. Let
the indices $j_S$ correspond to the two scalar Higgs bosons and let
$j_P$ correspond to the pseudoscalar Higgs. Then $\alpha_{j_S}=0$ or
$\pi$ and $\alpha_{j_P}=\pm \pi/2$ (i.e., the stops have real
(imaginary) couplings to the scalars (pseudoscalar) -- see
Eq.~(\ref{eq:alpha_def})). Furthermore,
$g^S_{\sss H_{j_P}\overline{\tau}\tau}=g^P_{\sss H_{j_S}\overline{\tau}\tau}=0$
(i.e., the Higgs bosons have pure scalar or pseudoscalar couplings to
the taus). It is clear, then, that Eq.~(\ref{eq:Asinglespin_approx})
is zero in the CP-invariant limit. Either the product of the scalar
and pseudoscalar couplings is zero, or the weak-phase difference is
$\pm\pi/2$, so that the cosine is zero.

Now suppose we allow CP to be broken by a small amount, so that there
are two ``mostly scalar'' and one ``mostly pseudoscalar'' Higgs bosons.
In this case there are two distinct types of contributions to the 
asymmetry. One contribution comes from the interference of 
a scalar and a pseudoscalar, denoted by the indices
$k$ and $j$, respectively. In this case we can redefine
$\widetilde{\alpha}_{j}=\alpha_j\mp \pi/2$ for the (mostly) 
pseudoscalar Higgs, so that $\widetilde{\alpha}_{j}$ measures
the departure of $\alpha_j$ from the CP-invariant limit of $\pm\pi/2$.
Then $\cos(\alpha_k-\alpha_j)=\pm\sin(\alpha_k-\widetilde{\alpha}_j)$,
and the expected sine of the weak phase difference appears.
Alternatively, suppose $j$ and $k$ are both mostly scalars,
but $g^P_{\sss H_j\overline{\tau}\tau}$ is not quite zero.
In this case there can be a contribution to the asymmetry
even if $\alpha_k=\alpha_j$. 

To summarize, the single-spin CP asymmetry can receive contributions
both from a weak phase difference in the interfering amplitudes, as
well as from scalar-pseudoscalar mixing of the Higgs bosons.  In
general there will be some combination of these effects.  Both types
of contributions require a strong phase difference.

\subsection{Triple-product Asymmetry}

Making the same approximations for the Higgs propagator as was made in
the previous subsection, we find the following expression in the case
of the triple-product asymmetry,
\begin{eqnarray}
      \left.\left|{\cal A}\right|^2\right|_{\mbox{\scriptsize TP}} 
      - \left.\left|\overline{\cal A}\right|^2\right|_{\mbox{\scriptsize TP}} 
      & \simeq & -8 M \sqrt{M^2-4 m_\tau^2} v^2 g_\tau^2 \nn \\
       & & \!\!\!\!\!\!\!\!\!\!\!\!\!\!\!\!\!\!\!\!\!\!\!\! \times 
      \left[\sum_k \frac{\left|
	  g_{\sss H_k\tilde{t}_2^*\tilde{t}_1}\right|^2
	  g^S_{\sss H_k\overline{\tau}\tau}g^P_{\sss H_k\overline{\tau}\tau}}
	{\left[\left(M^2-m_{\sss H_k}^2\right)^2+\Gamma_k^2 m_{\sss H_k}^2\right]} 
	   \right. \nn\\
	& & \!\!\!\!\!\!\!\!\!\!\!\!\!\!\!\!\!\!\!\!\!\!\!\! \left. +
	\sum_{j\neq k}
	  \frac{\left|
	  g_{\sss H_j\tilde{t}_2^*\tilde{t}_1}g_{\sss H_k\tilde{t}_2^*\tilde{t}_1}\right|
	  g^S_{\sss H_k\overline{\tau}\tau}g^P_{\sss H_j\overline{\tau}\tau}
          \cos\left(\alpha_k-\alpha_j\right)\cos\left(\delta_k-\delta_j\right)}
	  {\sqrt{\left[\left(M^2-m_{\sss H_k}^2\right)^2+\Gamma_k^2 m_{\sss H_k}^2\right]
	   \left[\left(M^2-m_{\sss H_j}^2\right)^2+\Gamma_j^2 m_{\sss H_j}^2\right]}}\right] .
	  \label{eq:TP_approx}
\end{eqnarray}
This expression bears some similarity to that derived for the
single-spin asymmetry (Eq.~(\ref{eq:Asinglespin_approx})), but with
two important differences. In the first place, the {\em cosine} of the
relative strong phase appears -- this asymmetry does not require a
strong phase difference. In the second place, there is a new term
compared to the single-spin asymmetry -- the term with a single sum
over $k$.  Since there is no need to have a relative strong phase for
the triple-product asymmetry, it is possible to receive contributions
from {\em single} resonances.  Such a contribution contains an
implicit dependence on CP-violating phases because it is only non-zero
if $g^S_{\sss H_k\overline{\tau}\tau}$ and $g^P_{\sss
H_k\overline{\tau}\tau}$ are both non-zero; i.e., if the Higgs boson
in question has both scalar and pseudoscalar couplings.  This can only
happen if CP has been broken.  Note that this first term is somewhat
analogous to the asymmetry considered by Valencia and Wang in
Ref.~\cite{ValWang}.



\begin{thebibliography}{99}
\bibitem{scenarios}
References~\cite{grossman} and
\cite{olive} contain nice discussions of several approaches that have
been considered.
\end{thebibliography}
\end{document}